\documentclass[preprint,review,12pt]{elsarticle}

\usepackage{lscape}
\usepackage{graphics}

\usepackage{amssymb}
\usepackage{wasysym}
\usepackage{multirow}

\journal{Nucl. Instrum. Meth. A}

\begin{document}

\begin{frontmatter}


   \cortext[cor1]{corresponding author. E-mail : otono@phys.kyushu-u.ac.jp.}

  \title{Development of time projection chamber for precise neutron lifetime measurement using
    pulsed cold neutron beams}


\author[KEK]{Y.~Arimoto}
\author[Tokyo]{N.~Higashi}
\author[KEK]{Y.~Igarashi}
\author[Kuicr]{Y.~Iwashita}
\author[KEK]{T.~Ino}
\author[Tokyo]{R.~Katayama}
\author[Kyoto]{R.~Kitahara}
\author[KMI]{M.~Kitaguchi}
\author[KEK]{H.~Matsumura}
\author[KEK]{K.~Mishima}
\author[Tokyo]{N.~Nagakura}
\author[Tokyo]{H.~Oide\fnref{CERN}}
\author[RCAPP]{H.~Otono \corref{cor1}}
\author[Nagoya]{R.~Sakakibara}
\author[RCNP]{T.~Shima}
\author[Nagoya]{H.~M.~Shimizu}
\author[Nagoya]{T.~Sugino}
\author[Kyushu]{N.~Sumi}
\author[Tokyo_Basic]{H.~Sumino}
\author[KEK]{K.~Taketani}
\author[Kyushu]{G.~Tanaka}
\author[KEK]{M.~Tanaka}
\author[KEK]{K.~Tauchi}
\author[KEK]{A.~Toyoda}
\author[Kyushu]{T.~Tomita}
\author[Tokyo]{T.~Yamada}
\author[ICEPP]{S.~Yamashita}
\author[Tokyo]{H.~Yokoyama}
\author[RCAPP]{T.~Yoshioka}

\fntext[CERN]{currently at CERN}

  \address[KEK]{High Energy Accelerator Research Organization, Ibaraki, Japan}
  \address[Tokyo]{Graduate School of Science, University of Tokyo, Tokyo, Japan}
  \address[Tokyo_Basic]{Department of Basic Science, University of Tokyo, Tokyo, Japan}
  \address[Kyoto]{Graduate School of Science, Kyoto University, Kyoto, Japan}
  \address[Kuicr]{Institute for Chemical Research, Kyoto University, Kyoto, Japan}
  \address[Kyushu]{Faculty of Sciences, Kyushu University, Fukuoka, Japan}
  \address[RCAPP]{Research Centre for Advanced Particle Physics, Kyushu University, Fukuoka, Japan}	
  \address[RCNP]{Research Center for Nuclear Physics, Osaka University, Osaka, Japan}
  \address[Nagoya]{Department of Physics, Nagoya University, Aichi, Japan}
  \address[KMI]{Kobayashi-Maskawa Institute, Nagoya University, Aichi, Japan}
  \address[ICEPP]{International Center for Elementary Particle Physics, University of Tokyo, Tokyo, Japan}
  
\begin{abstract}

    A new time projection chamber (TPC) was developed for neutron lifetime measurement
    using a pulsed cold neutron spallation source at the Japan Proton Accelerator Research Complex (J-PARC).
    Managing considerable background events from natural sources and the beam radioactivity is a challenging aspect of this measurement. 
    To overcome this problem, the developed TPC has unprecedented features 
    such as the use of polyether-ether-ketone plates in the support structure and internal surfaces covered with
    $^6$Li-enriched tiles to absorb outlier neutrons.
    In this paper, the design and performance of the new TPC are reported in detail. 
\end{abstract}


\end{frontmatter}


\section{Introduction}
\label{Introduction}

Neutron lifetime is an important observable parameter used to determine $V_{ud}$ in the Cabibbo-Kobayashi-Maskawa (CKM) matrix, together with the $\beta$-asymmetry parameter in neutron decay.
Furthermore, it is used as a probe to test the Big Bang theory through primordial nucleosynthesis.
After the first observation of neutron decay in 1948, various neutron lifetime measurements
using nuclear reactors have been conducted.
Currently, two main measurement methods exist with a $8.4\pm2.2$ s discrepancy between their results: one requires the counting of surviving ultra-cold neutrons after storing (giving $879.6\pm0.8$ s \cite{Mampe,Serebrov,Pichlmaier,Arzumanov,Steyerl}) and 
the other requires the counting of trapped protons from neutron decay (giving $888.0\pm2.1$ s \cite{Byrne,NIST2013}).
Therefore, accurate measurements using different methods are necessary.

The proposed experiment in this paper firstly employs an accelerator as a neutron source \cite{Shimizu}, 
which is conceptually based on a measurement by Kossakowski {\it{et al.}} \cite{Kossakowski} using a reactor and a time projection chamber (TPC).
The result of this measurement was published in 1987 as $878 \pm27~\rm{(stat.)} \pm14~\rm{(sys.)}$ s.
In this method, the TPC is filled with $\rm{^4He}$, $\rm{CO_2}$, and a few ppm of $\rm{^3He}$ gas,
and detects electrons emitted from the neutron decay while simultaneously measuring the neutron flux by counting $\rm{^3He(n,p)^3H}$ reactions.
To keep the number density of the ${\rm ^{3}He}$, the TPC is housed inside a vacuum vessel,
which is filled with gas after vacuuming of the vessel and sealed during operation.

The incident neutron beams are shaped into short bunches with lengths of approximately half the TPC.
The fiducial time during which the neutron bunch is entirely inside the TPC can be defined,
which enables us to reduce the uncertainty related to comparison between the number of the neutron decays and the number of the $\rm{^3He(n,p)^3H}$ reactions.

\section{Experimental overview}\label{sec:overview}

In this section, our experiment is described in comparison with the experiment conducted by Kossakowski {\it{et al.}}
The specifications of the beam and the TPC are summarized in Table \ref{tb:Beam}.
In particular, the TPC developed in this paper had two features which can be found in Table \ref{tbl:radioactivity_materials} and \ref{tb:LiF_gamma} in Section \ref{sec:TPCmaterial}:
\begin{itemize}
\item made of PEEK without radioactive contamination,
\item lined by $\rm{^6Li}$ tiles for the capture of scattered neutron.
\end{itemize}
Our TPC achieved the drift velocity of $1.0 \rm{cm/\mu s}$ under 100 kPa.
The multiplication gain was $4\times10^4$ with a resolution defined by FWHM of 22.9\% for 5.9 keV,
which resulted in 0.2 keV energy threshold per wire.
The detailed performance is described in Section \ref{sec:PerformanceTPC}.

The coordinate system is defined as follows: the $z$-axis is parallel to the direction of the neutron beam in the TPC, 
the $y$-axis is vertically aligned from the bottom to the top, and the $x$-axis is defined using the right-handed Cartesian coordinate system.

\subsection{The Kossakowski {\it{et al.}} neutron lifetime experiment}
\label{sec:Kossakowski}
Kossakowski {\it{et al.}} employed a continuous cold neutron beam from a reactor at the Institut Laue-Langevin (ILL), with a TPC of volume 190 ($x$) $\times$ 190 ($y$) $\times$ 700 ($z$) mm$^3$ \cite{Grivot}.
The neutron beam was shaped into bunches with lengths of 23--25 cm by a chopper drum rotating at 110 Hz, 
and monochromatized at a wavelength of 4.73 {\AA} by Bragg reflection on a graphite crystal \cite{Bussiere}.
The size and the divergence of the neutron beam were $15\times 25~\rm{mm^2}$ and $\pm 8.7~\rm{mrad}$, respectively.
The neutron flux inside the TPC was $2.2\times10^5$/s, which corresponded to 0.10 neutron decays/s for the fiducial time of 400 $\mu$s. 
A duty factor was calculated as 0.044.

The TPC consisted of a drift cage with a multi-wire proportional chamber (MWPC) inside the vacuum vessel.
The total gas pressure was fixed at 95 kPa, and a mixture of 93\% ${\rm{^4He}}$, 7\% $\rm{CO_2}$ and 0.7 ppm ${\rm{^3He}}$ was adopted.
The MWPC had sense/field wires in the $z$-direction sandwiched by two layers of cathode wires in the $x$-direction.
The wire cell in the MWPC was 10 $\times$ 10 mm$^2$.
The data acquisition was triggered by any hit on the sense wires.

Electrons from the neutron decays have a continuous kinetic energy spectrum up to 782 keV and deposit a part of this energy in the TPC;
The energy loss for the electrons with kinetic energy of $O(100)$ keV is less than 1 keV/cm.
On the other hand, the $\rm{^3He(n,p)^3H}$ reactions release monochromatic $Q$-value energy of 762 keV, and its decay products are both stopped inside the TPC.
Due to the saturation of the multiplication at the sense wires, as described in \ref{sec:SaturationModel},
the energy spectrum of the $\rm{^3He(n,p)^3H}$ reaction becomes broad and overlaps that of the neutron decay.
Kossakowski {\it{et al.}} used the maximum pulse amplitude among the sense wires as a discriminant variable and set 120 keV for the energy threshold.
An uncertainty of 0.6\% was assigned for the separation.

The neutrons are absorbed by the carbon in the gas at a rate comparable to the neutron decay.
The $\rm{^{12}C(n,\gamma)^{13}C}$ reaction generates a point-like energy deposit of 1.0 keV along the beam axis due to the $\rm{^{13}C}$ recoil by prompt $\gamma$-ray of 5.0 MeV.
To prevent these events, more than two sense wire hits were required.
Thus, $\epsilon_{carbon}$ represents the fraction of the loss due to the removal of the $\rm{^{12}C(n,\gamma)^{13}C}$ reactions, which was calculated as $11\pm1$\% in the Kossakowski {\it{et al.}} experiment.

In this approach, the triggered event data were divided into two subsets: $N_{\rm{^3He}}$ for the $\rm{^3He(n,p)^3H}$ reactions and $N_\beta$ for the neutron decays. 
The detection efficiency for the neutron decay, $\epsilon_{\beta}$, was more than 99.9\%, so that the uncertainty on $\epsilon_{\beta}$ was less than 0.1\%.
The neutron lifetime, $\tau_n$, is expressed as 
\begin{equation}
\tau_n = \frac{N_{\rm{^3He}}}{N_\beta-B_{env}-B_{beam}} \times  \frac{\epsilon_{\beta} \times(1-\epsilon_{backscatter}) \times (1-\epsilon_{carbon})}{\rho_{\rm{^3He}}\sigma_{\rm{^3He}}v_n},
\label{eq:tau}
\end{equation}
where $\rho_{\rm{^3He}}$, $\sigma_{\rm{^3He}}$, and $v_n$ are the number density of $\rm{^3He}$,
the cross section of the $\rm{^3He(n,p)^3H}$ reaction, and the neutron velocity, respectively.
During filling the gas, the value of $\rho_{\rm{^3He}}$ was determined by a pressure gauge with an uncertainty of 0.5\%, while 
$\sigma_{\rm{^3He}}v_n$ was treated as a constant using the $1/v$ dependence of $\sigma_{\rm{^3He}}$ for the cold neutron.
At $v_n=2,200$ m/s, $\sigma_{\rm{^3He}}$ was obtained as $5,333\pm7~\rm{barn}$ \cite{3He_1}. The remaining terms are described below.

Neutron decay events were identified only by detecting electrons by the TPC,
thus, there are many types of natural environmental background events labeled $B_{env}$: 
cosmic rays, $\gamma$-rays from natural radioisotopes, and radioactivity due to the detector materials.
Their TPC with cosmic-ray veto counters and lead shields had $B_{env}$ of 80 cps,
since no special low-activity materials were used.

In Eq.(\ref{eq:tau}) above, $N_\beta$ includes background events labeled $B_{beam}$ due to prompt $\gamma$-rays induced by the neutron beam, 
which occured at 20 cps.
The $\gamma$-rays from the upstream side of the TPC were prevented by the lead shields.
On the other hand, once the neutrons passing through the TPC were scattered by the gas and absorbed in the TPC itself or the vacuum vessel, 
they also became a $\gamma$-ray source.
The TPC was surrounded by a $^6$LiF coated plexiglas box,
since $\rm{^6Li}$ has a large neutron absorption cross section but does not emit prompt $\gamma$-rays.
However, neutron absorption by the TPC structure itself was not avoided in this setup.

In order to reduce this background events, 
one of the terminal points of the reconstructed tracks was required to be located on the neutron beam axis,
since the electrons emitted by the neutron decays originate there.
This requirement obstructed some of the electrons originating from the neutron decay which were scattered on the inner surface of the TPC.
The term, $\epsilon_{backscatter}$, represents the fraction of the loss, which was estimated by Kossakowski {\it{et al.}} as $1.5\pm0.5$\%.

Also, the large $B_{env}$ limited the statistical uncertainty to 3.1\%. 
Kossakowski {\it{et al.}} then estimated the value of $B_{beam}$ as 0.9\% of the neutron decay rate
and added the same amount of systematic uncertainty to the subtraction of $B_{beam}$ from the result.
All the uncertainties for the measurement are summarised in Table \ref{tb:unc}.

\subsection{Neutron beam for J-PARC experiment}

\begin{figure*}[tbp]
  \begin{center}
   \includegraphics[width=0.95\textwidth]{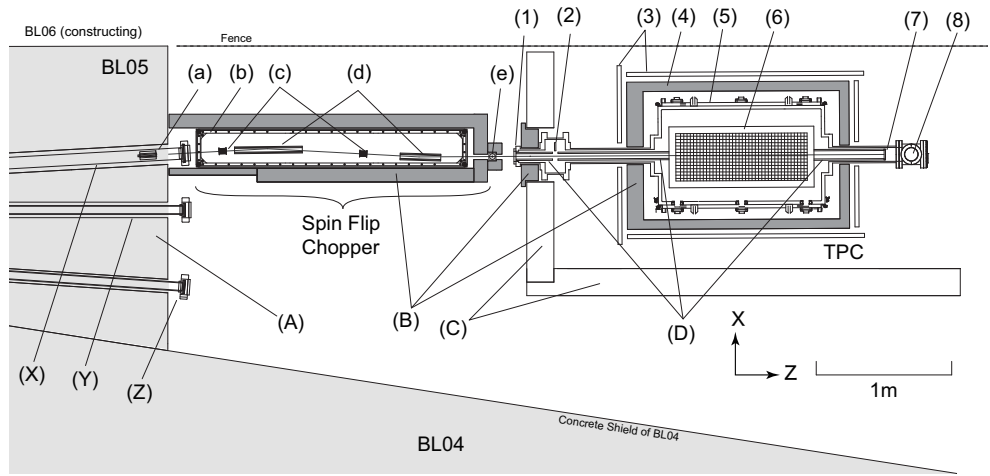}
  \end{center}
  \caption{Experimental setup at BL05 in MLF (top view).
    (A) BL05 concrete shield, (B) lead
    shield, (C) iron
    shield, (D) $^6$Li beam collimator, (X) polarized-beam branch, (Y)
    unpolarized-beam branch, (Z) low-divergence-beam branch, (a) short-wavelength pass filter, (b) guide coil,
    (c) resonance spin flipper coils, (d) magnetic supermirrors, (e) neutron beam monitor,
    (1) 50-$\rm{\mu m}$-thick Zr window, (2) neutron switching shutter, (3) cosmic-ray veto counter, (4) lead shield, (5) vacuum
    vessel, (6) TPC,
    (7) $^6$Li beam catcher, and (8) turbo
    molecular pump.}
  \label{fig:ExperimentalArea}
\end{figure*}

The Materials and Life Science Facility (MLF) at the Japan Proton Accelerator Research Complex (J-PARC) has one of the most intense pulsed neutron sources currently available.
3 GeV proton beams with a repetition rate of 25 Hz are injected into a liquid mercury target, 
which is supported from the downstream of the proton beam line.
The spalled neutrons emitted from the target are cooled by liquid hydrogen moderators
located above and below the target.
Our experimental setup, located at the polarized-beam branch of the neutron optics and physics (NOP) beamline BL05 port \cite{BL05}, is shown in Figure \ref{fig:ExperimentalArea}.

The polarized pulsed neutron beam has a broad time-of-flight (TOF) structure reflecting Boltzmann distribution.
The polarized beam is transported into the TPC via a neutron optical system of magnetic supermirrors and spin-flipper coils called 
a spin flip chopper (SFC) \cite{SFC}.
Resonance spin flipper coils can rotate the spin orientation of the neutrons, and 
the flipped neutrons then pass through the supermirrors and are dumped before the TPC.
By turning on the coils during the un-desired time domain, the beam is shaped into arbitrary-sized bunches.
The neutron flux is monitored using a thin $\rm{^3He}$ beam monitor \cite{BeamMonitor} at the end of the SFC.
The beam size is finally defined as $20 \times 20~\rm{mm^2}$ at the neutron switching shutter positioned between the SFC and the TPC.

Assuming a proton-beam power of 300 kW at J-PARC, the number of neutrons at the exit of the polarized-beam branch was calculated to be
$2.9\times10^7 /\rm{s}$ with an aperture of $30 \times 30~\rm{mm^2}$.
The divergence of the beam was $\pm4.2~\rm{mrad}$.
If the entire TOF region passed the SFC, $1.2\times10^6 /\rm{s}$ neutrons were guided inside the TPC.
When the SFC reflected five 40-cm-long bunches per pulsed beam, $1.7\times10^5 /\rm{s}$ neutrons were obtained.
In this configuration, the duty factor was calculated as 0.059.

\subsection{TPC for the J-PARC experiment}
\label{sec:WhatAreNews}
\begin{figure}[tbp]
  \begin{center}
    \includegraphics[width=0.8\textwidth]{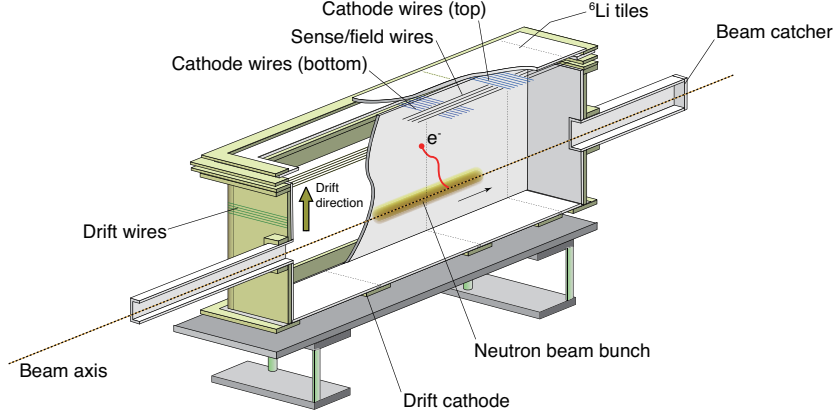}
  \end{center}
  \caption{Schematic view of our TPC. The TPC support structure is composed of PEEK plates.
  The $\rm{^6Li}$ tiles are set inside the drift cage and above the MWPC.
  A neutron bunch have a length of approximately half that of the TPC.}
  \label{fig:PEEKTPC_3D}
\end{figure}

The vacuum vessel was made from a stainless steel housing with internal size of $630~(x) \times 680~(y) \times 1180~(z)~{\rm mm}^3$ and three aluminum lids.
The TPC comprised a MWPC on a drift cage as shown in Figure \ref{fig:PEEKTPC_3D}.
The volume of the drift cage was $290~(x) \times 295~(y) \times 960~(z)~{\rm mm}^3$, and 0.092 neutron decays/s were expected in the fiducial time during 
which the neutron bunch was entirely inside a region of $\pm400$ mm at the center of the TPC.
The wire structure of the MWPC was the same as that of the TPC developed by Kossakowski {\it{et al.}}, and the wire cell size in the MWPC was 12 $\times$ 12 mm$^2$.

The expected neutron decay rate of 0.092 cps in the fiducial time was equivalent to the continuous events of 1.6 cps.
For further precision regarding the statistical uncertainty, reduction of $B_{env}$ was crucial,
especially in relation to the background events due to radioisotopes contained in the TPC system. 
We investigated several candidates for the TPC material and decided to use PEEK for the main structure as described in Section \ref{sec:PEEK}.

In order to reduce the systematic uncertainty concerning the subtraction of $B_{beam}$,
we developed a new tile containing $^6$Li as shown in Section \ref{sec:LiF} and covered the entire inner surface of the TPC with these tiles.
The pressure of the gas inside the TPC can be varied from 50 to 100 kPa in order to evaluate $B_{beam}$ using a data-driven method. 
If we use different pressures for the measurement, $p_1$ and $p_2$, the neutron lifetime can be expressed by Eq.(\ref{eq:tau});  
\begin{eqnarray}
\tau_n &=& \frac{N_{\rm{^3He}}(p_1)}{N_\beta(p_1)-B_{env}(p_1)-B_{beam}(p_1)} \times  \frac{\epsilon(p_1) }{\rho_{\rm{^3He}}\sigma_{\rm{^3He}}v_n} \nonumber \\
           &=& \frac{N_{\rm{^3He}}(p_2)}{N_\beta(p_2)-B_{env}(p_2)-B_{beam}(p_2)} \times  \frac{\epsilon(p_2) }{\rho_{\rm{^3He}}\sigma_{\rm{^3He}}v_n},
           \label{eq:tau1}
\end{eqnarray}
where $\epsilon$ denotes $\epsilon_{\beta} \times(1-\epsilon_{backscatter}) \times (1-\epsilon_{carbon})$ for simplicity.
Since $B_{beam}$ is proportional to the pressure, we have
\begin{equation}
\frac{B_{beam}(p_1)}{B_{beam}(p_2)}=\frac{p_1}{p_2}.
\label{eq:tau2}
\end{equation}
From Eq.(\ref{eq:tau1}) and Eq.(\ref{eq:tau2}), the neutron lifetime can be calculated without $B_{beam}$ as
\begin{equation}
\tau_n = \frac{p_1 \epsilon(p_2)N_{\rm{^3He}}(p_2) - p_2 \epsilon(p_1)N_{\rm{^3He}}(p_1) }{p_1(N_\beta(p_2)-B_{env}(p_2)) - p_2(N_\beta(p_1)-B_{env}(p_1))} \times \frac{1}{\rho_{\rm{^3He}}\sigma_{\rm{^3He}}v_n},
\label{eq:tau3}
\end{equation}
which would reduce the uncertainty related to $B_{beam}$ .
In this study, the gas mixture was fixed at ${\rm He/CO_{2}=85/15}$.
The $\rm{^3He}$ abundance was controlled as about $1~\rm{ppm}$.

Reduction of the systematic uncertainty regarding the correction for $B_{carbon}$ is possible by introducing
an energy threshold to distinguish the $\rm{^{12}C(n,\gamma)^{13}C}$ reaction from the neutron decay,
instead of just requiring more than two sense wire hits.
99.9\% of electrons from the neutron decay have kinematic energy of more than 4.0 keV.
If an energy threshold of 4.0 keV would be applied, the $B_{carbon}$ correction would be reduced to 0.1\%,
so that the uncertainty on this correction would also be reduced to less than 0.1\%.
The detailed design and performance of the TPC are discussed in Sections \ref{sec:TPCsystem} and \ref{sec:PerformanceTPC}, respectively.

\begin{table*}[htbp]
\begin{center}
\begin{tabular}{cccc}
\hline
&            & Kossakowski {\it{et al.}}  & This work\\
\hline\hline
     &Facility                         &      ILL                  & J-PARC (300 kW)\\
     &Repetition rate [Hz]             &      110                  & 25  \\
     &Pulse per repetition             &       1                   & 5\\
     &Beam size [mm$^2$]               &  15 $\times$ 25            & 20 $\times$ 20  \\
Beam &Pulse length [cm]                &  23--25                   & 40  \\
     &Beam divergence [mrad]           &  $\pm 8.7$                &  $\pm 4.2$ \\
     &Velocity [m/s]                   &      837                  & 500--1200   \\  
     &Duty factor for fiducial time (=$F$) &      0.044            & 0.059             \\ 
     &Neutron flux inside the TPC [1/s]  &     $2.2\times 10^5$  & $1.7 \times 10^5$             \\ 
     &Neutron decay rate [1/($F\cdot$s)]&        0.10        & 0.092  \\
\hline
    &Drift cage size [mm$^3$]           & 190 $\times$ 190 $\times$ 700 & 290 $\times$ 295 $\times$ 960 \\
	&MWPC cell size [mm$^2$]            & 10 $\times$ 10                & 12 $\times$ 12        \\
TPC &Gas pressure [kPa]                 & 95                            & 50 $\sim$ 100  \\
    &Gas mixture ratio [$\rm{He:CO_2}$] & $\rm{93:7}$                   & $\rm{85:15}$ \\
    &$\rm{^3He}$ abundance [ppm]        & 0.7                           & $\sim$ 1            \\
\hline
\end{tabular}
\end{center}
\caption{
Comparison of neutron beams and TPCs between the experiment by Kossakowski {\it{et al.}} and the present experiment, assuming 300-kW beam power at J-PARC.
}
\label{tb:Beam}
\end{table*}

\section{TPC material selection}\label{sec:TPCmaterial}

The PEEK and the $^6$Li tiles are the unique materials used for the TPC design presented in this paper.
In this section, their specifications are described in detail.

\subsection{PEEK as the TPC support structure}
\label{sec:PEEK}

Activity concentrations of the radioisotopes in the materials were determined by $\gamma$-ray spectrometry using a germanium detector in KEK, which was previously calibrated by Canberra \cite{LabSOCS_0} and shielded from environmental radiations by a 0.5-cm-thick acrylic plate, 0.5-cm-thick oxygen-free copper, and 10-cm-thick lead. 
The $\gamma$-ray count rate from the sample material was obtained by subtracting the count rate measured without the material.
The detection efficiency of each $\gamma$-ray was determined by using Canberra's LabSOCS software \cite{LabSOCS_1}. The nuclear data obtained from the Table of Isotopes \cite{TOI} were used for data analysis.

Table \ref{tbl:radioactivity_materials} shows the properties of Polyether-ether-ketone (PEEK) and Poly-phenylene-sulfide (PPS) produced by Yasojima Proceed Co.~Ltd with product names PEEK450G and FORTRON,
which had the first and the second least activity concentration emitting $\gamma$-rays among the candidates for the TPC support structure.
The PEEK possessed enough mechanical properties in terms of the wire tension, soldering and vacuuming.
Table \ref{tbl:radioactivity_materials} also shows the activity concentration of the $\rm{^6Li}$ tile described in Section \ref{sec:LiF},
stainless steel (SUS304) and aluminum (A5052) for the vacuum vessel provided by VIC International,~Inc.

The dominant radionuclides inducing the background events in the TPC were $^{210}$Pb and its daughters,  $^{210}$Bi and  $^{210}$Po,
according to the $\gamma$-ray spectrometry. 
The radiations ($\alpha$-rays, $\beta$-rays, $\gamma$-rays, and bremsstrahlung) emitted following the decay of $^{210}$Pb and its daughters
in the PEEK and the SUS304 were simulated by the {\tt{Geant4}} \cite{Geant4}
with the measured activity concentrations and the detailed geometry of the detector system as described in Section \ref{sec:TPCsystem}.
The activities of the daughter radionuclides were assumed to be in equilibrium with $^{210}$Pb. 
The trigger rates of the TPC induced by the radiations from $^{210}$Pb and its daughters were anticipated to be $2.2\pm1.5$ cps from the TPC support structure and $0.2\pm0.1$ cps from the vacuum vessel.
The uncertainty was derived from the measurements of the activity concentration by the germanium detector.
In total, the expected background rate was $2.4\pm1.5$ cps.
Comparison with the observation can be shown in Section \ref{sec:BackgroundTPC}.

\begin{table*}[htbp]
\centering
\begin{tabular}{cccccc}
\hline
   &Total $\gamma$-radio-    &  \multirow{2}{*}{$^{210}$Pb}  &Elastic              & Melting                   & Water \\
   &nuclides                           &                                               & modulus           &  point                     & absorption\\
   & [${\rm Bq/cm^{3}}$]        &[${\rm Bq/cm^{3}}$]                 & [${\rm GPa}$]  &  [${\rm ^{\circ}C}$]  &  [\%]   \\
 \hline\hline
PEEK               &  $0.017\pm0.012$  & $0.022\pm0.011$  &  3.6       &334&0.14\\
PPS                 &  $0.057\pm0.015$  & $0.053\pm0.015$  & 3.9        &278&0.04\\
$\rm{^6Li}$ tile &  $-0.011\pm0.022$ & $-0.010\pm0.022$ &---            &---&---                          \\
SUS304           &  $1.5\pm1.0$          & $1.5\pm1.0$          & ---          &---&---\\
A5052              &  $-0.019\pm0.080$ & $-0.028\pm0.080$ &---            &---&---                          \\
\hline
\end{tabular}
\caption{
Total activity concentration of the radionuclides emitting $\gamma$-rays and activity concentration of $^{210}$Pb in the materials for the TPC and the vacuum vessel, determined by $\gamma$-ray spectrometry.
Mechanical properties of PEEK and PPS are taken from the specification sheet of PEEK450 and FORTRON produced by Yasojima Proceed Co.~Ltd.}
\label{tbl:radioactivity_materials}
\end{table*}

\subsection{$^{6}Li$ tiles inside the TPC}\label{sec:LiF}

To produce the $\rm ^{6}Li$ tile, ${\rm ^{6}Li}$-enriched lithium fluoride (LiF) powder was baked with poly-tetra-fluoro-ethylene (PTFE).
The ${\rm {}^{6}Li}$ had a ${\rm 95~\%}$ concentration in the lithium. 
The mixture ratio of the LiF was 30\% by weight due to the stiffness property.
The thickness of the $^6$Li tile was 5 mm, and its molding size before manufacturing was 300 $\times$ 300 mm$^2$.
The absorption length of the $^6$Li tile was 0.5 mm for the thermal neutrons, so that the neutrons scattered by the gas can be completely absorbed. 
As Table \ref{tb:LiF_gamma} shows, the ratio of the neutron absorption with the prompt $\gamma$-rays in the $\rm ^{6}Li$ tile was reduced to $8.3\times10^{-5}$ calculated by using cross sections in ref. \cite{NIST}.
If the prompt $\gamma$-rays are emitted, the average number of the prompt $\gamma$-rays was calculated as 2.0,
which was based on the NNDC on-line data service from the ENSDF database \cite{NNDC}.
In the developed TPC, 100-$\rm{\mu m}$-thick sheets of PTFE were attached to the $^6$Li tile in order to prevent $\alpha$ and ${\rm {}^{3}H}$ particles from returning to the TPC active volume. 

\begin{table}[htb]
\begin{center}
\begin{tabular}{ccccc}
\hline
                 &Absorption cross                    & Mole            &Branching              &average number \\
                 &section with $\gamma$-ray    & fraction        &  ratio                     &of $\gamma$-ray\\
\hline\hline
$\rm{^6Li}$      & 39 mbarn                        & 0.17           & $4.1\times10^{-5}$ & 1.4 \\
$\rm{^7Li}$      & 45 mbarn                       & 0.01            & $2.5\times10^{-6}$ & 1.0\\
$\rm{C}$          & 3.5 mbarn                      & 0.21           & $4.5\times10^{-6}$ & 1.2 \\
$\rm{F}$           & 39 mbarn                       & 0.61           & $3.5\times10^{-5}$ & 2.8 \\
\hline
$\rm{^6Li}$ tile & 32 mbarn                       & 1.0             & $8.3\times10^{-5}$ & 2.0\\
\hline
\end{tabular}
\end{center}
\caption{Properties of the prompt $\gamma$-rays induced by neutron capture in the $\rm{^6Li}$ tile.}
\label{tb:LiF_gamma}
\end{table}

\section{The detector system}
\label{sec:TPCsystem}

\begin{figure}[tbp]
  \begin{center}
    \includegraphics[width=0.65\textwidth]{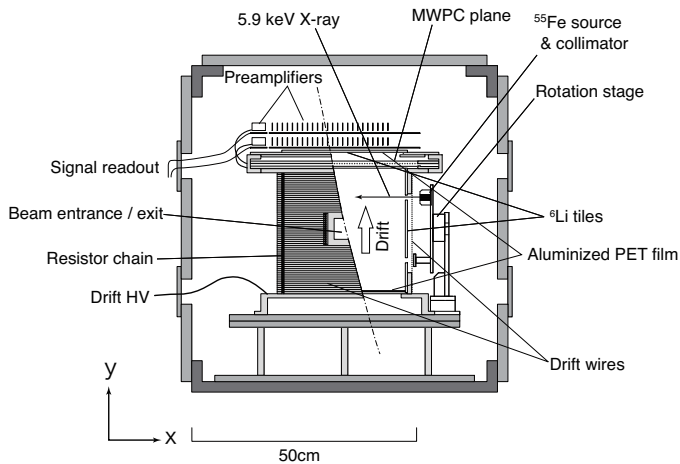}
  \end{center}
  \caption{TPC in vacuum vessel viewed from downstream side (x-y plane).}
  \label{fig:PEEKTPC_FRONT}
\end{figure}

In this section, the structure of the drift cage and the MWPC in our TPC are explained in detail.
Figure \ref{fig:PEEKTPC_FRONT} shows a cross section of the TPC.

\subsection{Drift cage}
\label{sec:driftcage}

The support structure of the drift cage was composed of PEEK plates which were welded into frames for the top and base sides,
horizontal planks for the longer side, and perpendicular panels with a rectangular neutron-beam-passing hole for the shorter side.
At the corners of the cage, 5-mm-pitch drift wires ($\rm \diameter50$-$\mu m$ of BeCu alloy) were soldered on a polyimide sheet with an etched electric circuit pattern.
2-${\rm M\Omega}$ chip resistors interconnected the drift wires.

The $\rm{^6Li}$ tiles were fixed inside the PEEK structure.
The base and longer side surfaces were covered with three and half $\rm{^6Li}$ tiles,
and the shorter side surfaces were covered with one $\rm{^6Li}$ tile. 
The edges of the $\rm{^6Li}$ tiles were milled so that the adjacent $\rm{^6Li}$ tiles overlaped by 5 mm in order to close any gaps. 
A 12-${\rm{\mu m}}$-thick aluminized PET film was added to the surface of the bottom $\rm{^6Li}$ tiles, for the application of a negative voltage for drifting of the ionized electrons. 

The presence of dielectric materials such as the $\rm{^6Li}$ tiles and the PEEK structure inside the drift wires could distort the drift field, and this effect must be considered in the design of the drift cage. 
The relative permittivities of the $\rm{^6Li}$ tile and the PEEK structure were measured as 3.0 and 3.2, respectively. 
We then simulated the drift field using a three-dimensional static electric field analysis to make the equipotential surface parallel to $xz$-plane. The calculated drift voltage was uniform within 1.7\% in the cage.
Note that wire structures of MWPC and charging up of materials were not taken into account for this simulation. 

A $40\times40$-$\rm{mm}^2$ and a $60\times60$-$\rm{mm}^2$ apertures were placed at the entrance and exit of the cage, respectively, to allow the neutron beam to pass, taking the divergence of the beam into consideration.
The beam ducts were connected to them, and the $\rm{^6Li}$ tiles covered the inner surfaces of the ducts as well as the TPC.
As a beam catcher, a PTFE box filled with the ${\rm ^{6}Li}$-enriched LiF powder was located at the end of the beam duct.

\subsection{The MWPC}
\label{sec:MWPC}
The three-layer MWPC wire structure with 6-$\rm{mm}$ gaps was made on two 6-mm-thick PEEK frames.
The outer and inner sizes of the PEEK frames were $412~(x) \times 1,070~(z)~{\rm mm}^2$ and $312~(x) \times 970~(z)~{\rm mm}^2$, respectively.
A single frame was used to stretch both the sense/field wires in the $z$-direction and the top cathode wires in the $x$-direction; 
the other frame had the bottom cathode wires on one side.
An additional PEEK frame was mounted above the MWPC so that the top $\rm{^6Li}$ tiles could be fixed upon it.
The top $\rm{^6Li}$ tiles were covered with aluminized PET film, for the application of a positive voltage,
such that the ionized electrons above the MWPC inside the TPC were collected into the film.

In the MWPC anode layer, 28 sense wires ($\rm \diameter20$-$\mu m$ of gold-plated tungsten) and 27 field wires ($\rm \diameter50$-$\mu m$ of BeCu alloy) were alternatively placed at every 6-mm pitch in the $z$-direction.
Each cathode layer consisted of 160 6-mm-pitch cathode wires ($\rm \diameter50$-$\mu m$ of BeCu alloy) in the $x$-direction.
The wire positions were determined by 100-$\rm{\mu m}$-width gutters.

The signals from 24 sense wires and 24 field wires in the central region were individually read out. 
Four sequential cathode wires were linked to a single channel, which corresponded to 40 channels per cathode layer. 
In total, the TPC had 128 channels.

Charges induced on the MWPC wires were converted to the voltage signal by charge-sensitive amplifiers.
The amplifiers were mounted on the top of the TPC.
Two conversion factors of 1.0 V/pC (high gain) and 0.1 V/pC (low gain) were implemented.
The high-gain amplifiers were for the sense and bottom cathode wires, and the low-gain amplifiers were for the field and top cathode wires.
It was confirmed that the ${\rm ^{3}He(n,p)^{3}H}$ events can be recorded without exceeding the dynamic range,
because of the space-charge effects occurring during the multiplication.

\subsection{Data acquisition system}

A common pipelined platform for electronics readout (COPPER) developed at KEK \cite{DAQ} was employed for our experiment.
We used COPPER lite, which was a successor to the COPPER device.
A COPPER board can mount up to four digitization daughter cards, called front-end instrumentation entities for sub-detector specific electronics (FINESSE).

The signals from the 24 TPC sense wires were fed into discriminators with 20-mV threshold for the pulse height, and a logical sum of the 24 outputs from the discriminators triggered the data acquisition.
A TDC with a 1.25-MHz clock and 16-bit dynamic range was employed to measure the time period from the injection of the latest proton pulse on the mercury target and the generation of the trigger signal.
The waveforms of the 128 channels from the TPC were digitized using a flash-ADC-type {\scshape FINESSE} card with a 20-MHz clock, 12-bit dynamic range, and 512-words/channel FIFO.

\begin{figure}[tbp]
  \begin{center}
    \includegraphics[width=1.0\textwidth]{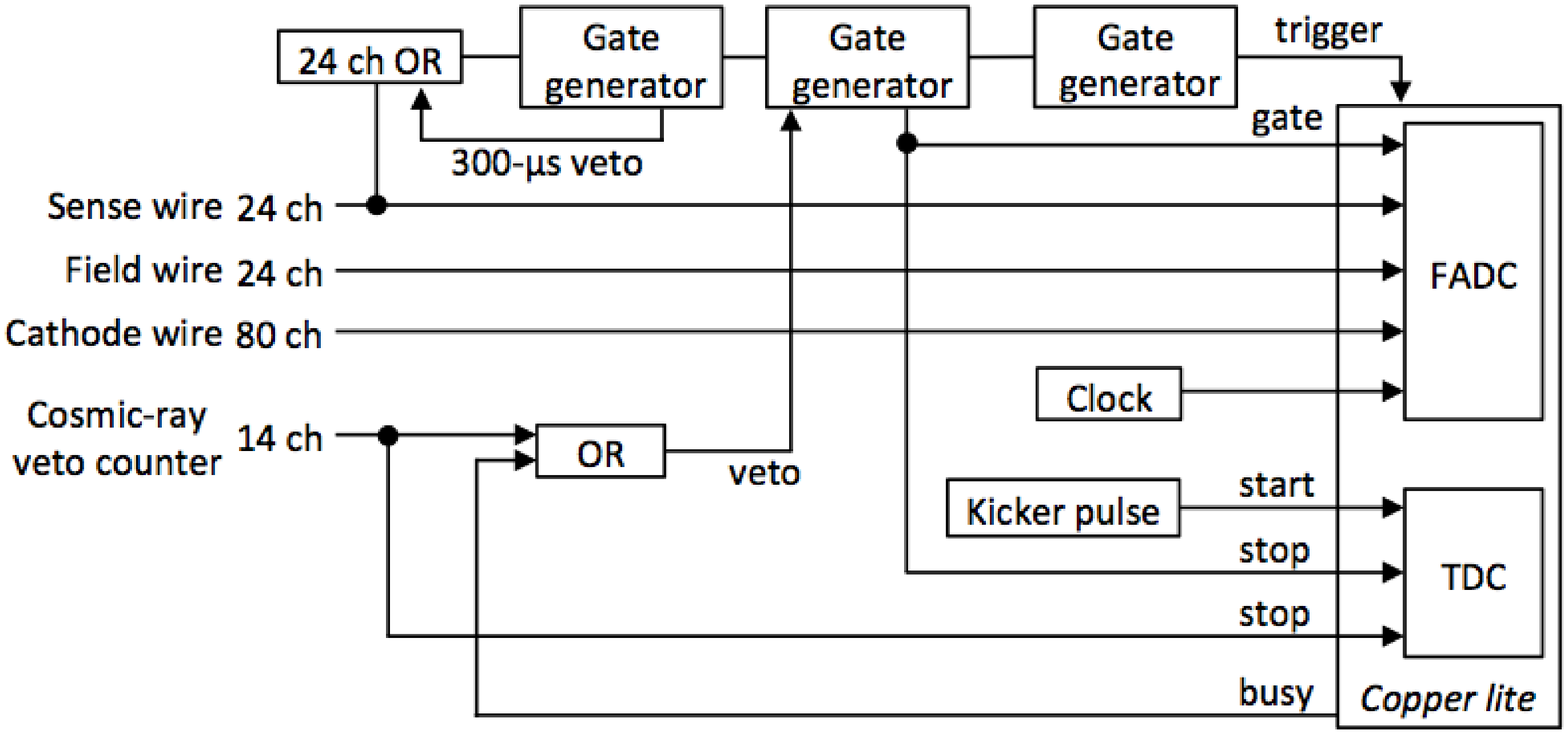}
  \end{center}
  \caption{Box diagram of the trigger and data acquisition.}
  \label{fig:DAQ}
\end{figure}

\subsection{Calibration system}

To monitor the TPC conditions, a ${\rm {}^{55}Fe}$ source with a collimator was equipped on a vertical rotation stage as shown in Figure \ref{fig:PEEKTPC_FRONT}.
Two slits at the side of the drift cage can be selected for the injection of 5.9 keV X-rays, with distances from the MWPC of 75 and 225 mm.
The X-rays passing through the collimator spreaded in the $xz$-plane in a trapezoidal shape, with a $y$-divergence of 10 mrad and covering 60\% of the MWPC area.
When the source was not used, the slits were closed by small $^6$Li tiles that were also located on the rotation stage. 

\subsection{Shields}\label{sec:Shielding}
The vacuum vessel was housed in a lead shield which covered all six sides except the beam path.
The lead was of 10-cm thickness on the upstream side, and 5-cm-thickness on the other five sides.  
The hermeticity of the lead shield was 99~\% because of the holes for the beam ducts.
The lead shield was further reinforced by a 20-cm-thick iron shield on the upstream side and on the side of the TPC that met another beam line, ``BL04''.
According to the NaI detector measurement, 98~\% of the environmental $\gamma$-rays were suppressed.

\subsection{Cosmic-ray veto counters}
The lead shield surfaces, apart from the bottom, were covered with cosmic-ray veto counters.
Both the lead shield and the cosmic-ray veto counters had square holes around the beam duct.
A panel of cosmic-ray veto counters consisted of a two-layer array of extruded plastic scintillator boards, with the layers being optically isolated.
The scintillator had trenches with 20-mm pitch, and a wavelength shifting optical fiber was inserted in each trench.
The fibers of the layers were bundled and read out by a photomultiplier tube.

The performance of the cosmic-ray veto counters was evaluated using a cubic scintillator with a side length of 10 cm,
which was also placed on the center of the bottom lead shield. 
The cosmic rays deposited approximately 20 MeV in the cubic scintillator, while other environmental background sources, including the
neutron-induced background, deposited at most 8 MeV. 
Thus, pure cosmic rays could be extracted. 
The measured efficiency of the cosmic-ray veto counters was found to be 96\%.

\subsection{Apparatus for determination of $\rm{^3He}$ partial pressure}

To determine the $\rm{^3He}$ partial pressure in the TPC, we used a pressure gauge produced by Mensor Corp. (Mensor Digital Pressure Gauge Series 2500),
which was equipped with a piezoresistive transducer. 
This gauge achieved 0.01\% accuracy for atmospheric pressure, as confirmed by NIST-traceable standards.
Current precision for the determination of the $\rm{^3He}$ abundance of $1~\rm{ppm}$ is 0.3\%.
We have also developed a sector-type single-focusing mass spectrometer in the
University of Tokyo \cite{MS} in order to directly measure the $\rm{^3He}$ and $\rm{^4He}$ ratio. 
The mass spectrometer employed a double collector system for simultaneous detection of $\rm{^3He}$ and $\rm{^4He}$.

\section{TPC performance}
\label{sec:PerformanceTPC}
In this section, we discuss in detail the detection system performance in terms of various TPC characteristics.

\subsection{Detector configuration}

For measurement of the drift velocity, the multiplication gain, and the detection efficiency, the top $^6$Li tiles on the TPC were removed. 
At that time, the MWPC was covered with ethylene vinyl alcohol (EVOH) film to obstruct ionized electrons from above the TPC, 
so that the particle response of the TPC would be unaffected.
For the evaluation of the long-term stability and background event rate, the top $^6$Li tiles on the MWPC were mounted in the normal setup.

According to Eq. (\ref{eq:tau3}), the measurements of the neutron lifetime under several gas pressure could reduce the systematic uncertainty on the subtraction of $B_{beam}$. We investigated the detection efficiency under 50 kPa and 100 kPa.
For the drift velocity and the multiplication gain, 75 kPa was also evaluated. 

\subsection{Drift velocity}\label{sec:drift}

Since the total count rate of the TPC was about 1 kcps which was dominated by the cosmic rays, 
the memory time of the TPC should be less than a few tens $\mu s$ in order to keep the dead time of the data acquisition below $O(1)$\%, 
and the drift velocity should be faster than about 1 cm/$\mu s$ with the maximum drift distance of 295 mm of the TPC. 

The drift velocities of the ionized electrons in the TPC, $v_{d}$, were measured using the cosmic rays which traversed the TPC in the $y$-direction. 
The timing difference, $\Delta t$, between the earliest and the latest hits on the sense wires was calculated.
$\Delta t_{\rm end}$ was defined as the end point in the distribution of $\Delta t$.
Due to the pitch of the anode wire, the apparent drift distance, $v_{d}\times \Delta t_{\rm end}$, was gradually close to 295~mm,
if the angle between the cosmic ray and the drift direction would be rectangular.
Thus we analyzed the cosmic rays with different angles and took the limit for the horizontal cosmic ray in order to obtain $v_{d}$. 
We implemented this method in the simulation, and checked an output of the calculation 
for a known velocity of the cosmic ray as an input.
Systematic uncertainty of $v_{d}$ was set as 4\%, which was determined by taking the maximum deviation between 
the inputs and the outputs of the simulation.

\begin{figure}[tbp]
  \begin{center}
    \includegraphics[width=0.50\textwidth]{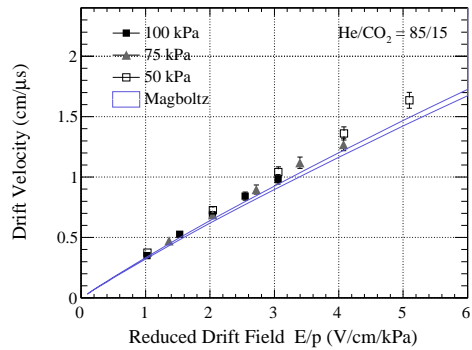}
  \end{center}
  \caption{Drift velocity dependence on reduced drift field and gas pressures, where $E$ and $p$ are the electric field and total pressure, respectively.
  The gas mixture was fixed at ${\rm He/CO_{2}=85/15}$.
  The band shows the calculated values with the Magboltz simulation including an uncertainty of temperature\cite{Magboltz}.}
  \label{fig:DriftVelocity}
\end{figure}

Figure \ref{fig:DriftVelocity} shows $v_{d}$ as a function of the reduced electric field for the gas configuration,
which is compared with the {\tt{Magboltz}} simulation \cite{Magboltz}. 
Temperature in the vacuum chamber varied from 298 K to 309 K during the measurements, 
thus we draw the results as a band calculated with the minimum and maximum temperatures.
Under 50 kPa and 100 kPa gas pressure for the other measurements in this paper, 
the high voltage for the drift cage was set as $-6,000$ V and $-9,000$ V to satisfy the requirement,
which corresponded to $v_{d}$ of 1.4 cm/$\mu s$ and 1.0 cm/$\mu s$, respectively. 
The measured velocities were slightly higher than Magboltz simulation, but could not affect the measurement of the
neutron lifetime, since we would use the measured velocities themselves.


\subsection{Multiplication gain}
\label{sec:gain}

\begin{figure}[htbp]
  \begin{center}
    \includegraphics[width=0.50\textwidth]{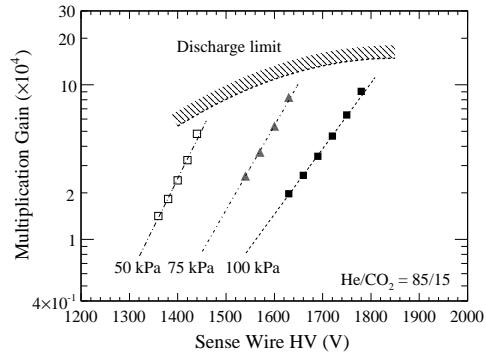}
  \end{center}
  \caption{Multiplication gain dependence of the MWPC on voltage for sense wires and total pressure values.
  The gas mixture was fixed at ${\rm He/CO_{2}=85/15}$.
  Discharge limit was empirically obtained.}
  \label{fig:Gain_GasDependence}
\end{figure}

The multiplication gain was measured using the energy deposit of 5.9 keV X-rays from the ${\rm ^{55}Fe}$ source.
The ${\rm ^{55}Fe}$ source was placed directly above the MWPC, such that the X-rays were directly absorbed in the MWPC wire-cell volume without drift. 
A $W$-value of 42 eV was used as an average energy to produce an ionized electron in the gas.
Figure \ref{fig:Gain_GasDependence} shows that our TPC achieved the multiplication gain of $4\times10^{4}$,
which allows us to apply the threshold of 0.2 keV by the signal-sensing threshold at 20 mV.
The high voltages for the sense wires under 50 and 100 kPa were set to be 1,440 and 1,750 V, respectively.

\begin{figure}[htbp]
  \begin{center}
    \includegraphics[width=0.50\textwidth]{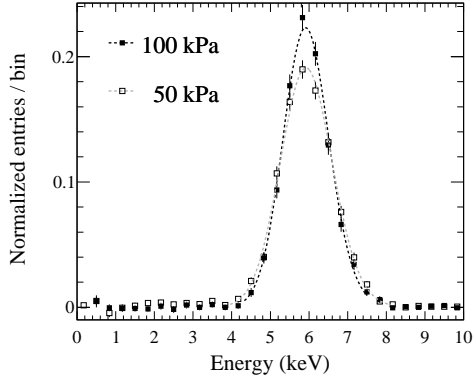}
  \end{center}
  \caption{
Pulse height spectra of $^{55}$Fe X-ray on a central anode wire taken with 1,400 V at 50 kPa and with 1,720 V at 100 kPa.
The resolutions in FHWM were 26.1\% and 22.9\% for 50 kPa and 100 kPa, respectively.
}
  \label{fig:FeSpectra}
\end{figure}

Pulse height spectra for 5.9 keV X-ray on a central wire can be shown in Figure \ref{fig:FeSpectra}. 
The background events were subtracted by rotating the ${\rm ^{55}Fe}$ source stage.
By fitting 2$\sigma$ regions of the peaks, the energy resolutions for 5.9 keV defined by the FWHM were 
obtained as 26.1\% with 50 kPa at 1,400 V and 22.9\% with 100 kPa at 1,720 V. These resolutions would realize sufficient separation between the neutron decay and the 1.0-keV $\rm{^{12}C(n,\gamma)^{13}C}$ reaction.
The reduction of the uncertainty on $\epsilon_{carbon}$ is discussed in Section \ref{sec:attenuation} together with the long-term stability.

\subsection{Detection efficiency for minimum ionizing particles}

The detection efficiency for the minimum ionizing particles with minimum path length on the sense wires was measured using the cosmic rays. 
At the normal position of the TPC, the required direction was achieved only by horizontal cosmic rays, but the numbers were not sufficiently high.
In order to use perpendicular cosmic rays instead, 
we rotated the TPC by $90^{\circ}$ around the beam axis in the vacuum vessel,
which has an almost square cross section (see Figure \ref{fig:PEEKTPC_FRONT}).
After the rotation, one side faced up and the other side turned to the bottom,
while the MWPC plane was perpendicular to a horizontal plane.
Two plastic scintillators for triggering were set both over and under the vacuum vessel. 
Cosmic rays with a zenith angle of less than $\pm 5^{\circ}$ were then selected.
The detection efficiency of the sense wire was determined by taking the hit ratio when both adjacent sense wires showed hits.

\begin{figure}[htbp]
  \begin{center}
    \includegraphics[width=0.50\textwidth]{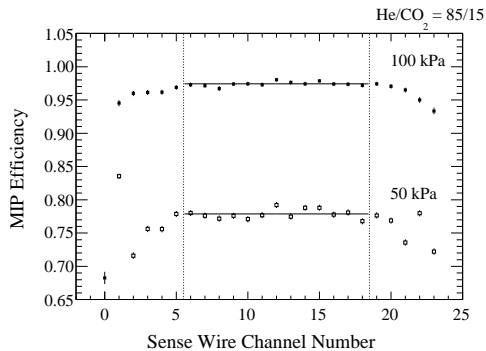}
  \end{center}
  \caption{Detection efficiency for the minimum ionisation particles with minimum path length on sense wires.
  The gas mixture was fixed at ${\rm He/CO_{2}=85/15}$.}
  \label{fig:Anode_DetectionEfficiency}
\end{figure}

The expected average energy deposit on one sense wire was found to be 0.5 and 1.0 keV for 50 and 100 kPa, respectively, 
while the hit discrimination threshold was 0.2 keV for both cases.
The results are presented in Figure \ref{fig:Anode_DetectionEfficiency}. 
The average efficiencies in the center of the TPC were 78 \% for 50 kPa and 97 \% for 100 kPa. 
Using the {\tt{Geant4}} and implementing these efficiencies, the detection efficiency for the neutron decay, $\epsilon_{\beta}$, under both total pressure values was estimated to be over 99.9\%, so that the uncertainty on $\epsilon_{\beta}$ should be less than 0.1\%.

\subsection{Long-term performance stability}
\label{sec:attenuation}

Since the gas is sealed during operation of the TPC, the attachment of the ionized electrons in the drift cage is caused by contaminations due to 
outgassing. 
The ${\rm ^{55}Fe}$ source on the rotation stage provides two drift lengths, which yields an attachment coefficient, $C~\rm{[m^{-1}]}$.
A large attachment coefficient means degradation of the energy resolution.

\begin{figure}[htbp]
  \begin{center}
    \includegraphics[width=0.50\textwidth]{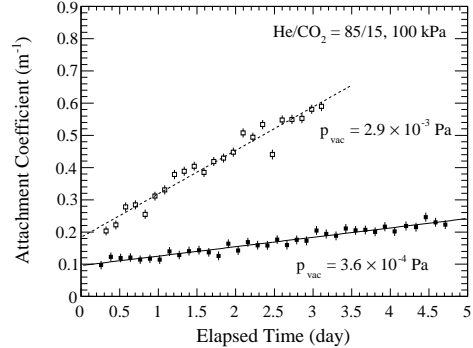}
  \end{center}
  \caption{Time variation of transportation efficiency of the TPC with different achieved pressures.
  The gas mixture was fixed at ${\rm He/CO_{2}=85/15}$.}
  \label{fig:Time_Attenuation}
\end{figure}

The outgassing rate was related to the pressure achieved during the evacuation before the gas filling, $P_{vac}$, and 
$3.6\times10^{-4}~{\rm Pa}$ was reached after an evacuation of approximately 2 weeks.
In this condition, the outgassing rate was 4.4 Pa/day.
Figure \ref{fig:Time_Attenuation} shows the long-term evolution of the attachment coefficient for the two different $P_{vac}$.
The TPC achieved $C < 0.25~{\rm m^{-1}}$ for 5 days of continuous operation,
which resulted in an energy threshold of 1.4 keV to reject the 1.0-keV $\rm{^{12}C(n,\gamma)^{13}C}$ reaction. 
With regard to $\epsilon_{carbon}$,
a 99.9\% efficiency for electrons from the neutron decay can be obtained with an energy threshold of 4.0 keV, 
so that the uncertainty of the $\epsilon_{carbon}$ correction should be less than 0.1\%.

The dielectric materials such as the $^6$Li tile were exposed inside the TPC, 
which may result in distortion of the electric field by charging.
During the 5-day operation, the deviation of the drift velocity and the multiplication gain were found to be 1\% and 10\%, respectively, 
which would not affect the neutron lifetime measurement, since the variation can be monitored and calibrated correctly. 

\subsection{Background event rate}
\label{sec:BackgroundTPC}

In the MLF, the value of $B_{env}$ without any rejection of the environmental $\gamma$-rays and cosmic rays was 123.7 cps. 
To evaluate the background events, the same voltage as in the detection efficiency measurement was applied.
$B_{env}$ was reduced to 58.4 cps by the lead shields.
Finally the cosmic ray counters reduced $B_{env}$ to 7.7 cps.
The count rates of the TPC due to the environmental $\gamma$-rays and cosmic rays would be 1.3 and 2.0 cps,
which were calculated from the performance of the lead shield and the cosmic-ray veto counters described in Section \ref{sec:TPCsystem}.
The remaining background rate of 4.4 cps would be due to the radioisotopes in the TPC and the vacuum vessel, 
which was almost consistent with $2.4\pm1.5$ cps of the expectation shown in Section \ref{sec:PEEK}.
With the improved $B_{env}$, a statistical uncertainty of 1.0\% can be achieved in 34 hours, which is the same data acquisition time as
that of Kossakowski {\it{et al.}}

Further, the $\rm{^6Li}$ tiles inside the TPC can reduce $B_{beam}$ to 2.0 cps.
Although the current systematic uncertainty on the subtraction of $B_{beam}$ has not yet been measured, it should be less than 0.9\% because of the significant reduction of $B_{beam}$.
All the uncertainties on the neutron lifetime measurement, including the improvements, are summarized in Table \ref{tb:unc}.

\begin{table*}[htbp]
\begin{center}
\begin{tabular}{ccc}
\hline  
                                                     & Kossakowski {\it{et al.}}    & This work \\
\hline\hline
     $B_{env}$                    & 80 cps                   &  7.7 cps  \\
     $B_{beam}$                   & 20 cps                   &  2.0 cps  \\
     $\epsilon_{carbon}$          & 11\%                   &  0.1\%  \\
\hline
     Unc. on statistics                                                                         &   3.1\%  &   1.0\% \\
     Unc. on separation between $N_{\rm{^3He}}$ and $N_\beta$   &   0.6\%  &   0.6\% \\
     Unc. on subtraction of $B_{beam}$                                            &   0.9\%  & $<0.9\%$ \\
     Unc. on correction of $\epsilon_{\beta}$                                     & $<0.1\%$ & $<0.1\%$ \\
     Unc. on correction of $\epsilon_{backscatter}$                           &   0.5\%  &   0.5\% \\
     Unc. on correction of $\epsilon_{carbon}$                                   &   1.0\%  & $<0.1\%$ \\
     Unc. on $\rho_{\rm{^3He}}$                                                         &   0.5\%  &   0.3\% \\
     Unc. on $\sigma_{\rm{^3He}}v_n$                                                &   0.1\%  &   0.1\% \\
\hline
     Total uncertainty                                           & 3.5\% & $<1.6\%$\\
\hline
\end{tabular}
\end{center}
\caption{
Comparison of $B_{env}$, $B_{beam}$, $\epsilon_{carbon}$ and neutron lifetime uncertainties  
between the experiment conducted by Kossakowski {\it{et al.}} and the present experiment, assuming 300-kW beam power at J-PARC and data acquisition time of 34 hours.
}
\label{tb:unc}
\end{table*}

\section{Conclusion}

A new measurement approach is being used at J-PARC in Japan, which has several unique advantages compared to the current measurement methods.
The TPC with PEEK plates and $\rm{^6Li}$ tiles was developed to enhance neutron decay detection during long term operation using a low background environment and improved signal efficiency.
Our experiment with this TPC will be able to reach 1.6\% of the precision on the neutron lifetime in 34 hours.
The expected systematic uncertainty is dominated by the subtraction of the background due to the neutron beam interacting with the gas inside the TPC,
which can be improved by the comparison of the measurements among the different pressures of the gas.
We have proved the TPC to keep the performance under the pressure from 50 kPa to 100 kPa, so that further precision on the neutron lifetime would be possible.

\section*{Acknowledgement}
We would like to thank Takashi Kobota for his effort in the early stage of this experiment.
We wish to express our gratitude to Setsuo Sato for his assistance to design of the preamplifier.
This research was partially supported by the Ministry of Education, Science, Sports and Culture, Creative Scientific Research 19GS0210, Scientific Research (A) 23244047, Scientific Research (B) 20340051, JSPS Fellows 256088, and JSPS Fellows 15J04202. 
The neutron scattering experiment was approved by the Neutron Scattering Program Advisory Committee of IMSS, KEK (Proposal No. 2009S03 and 2014S03). The neutron scattering experiment was approved by the Neutron Science Proposal Review Committee of J-PARC/MLF (2012A0075, 2012B0219, 2014A0244, and 2014B0271) and supported by the Inter-University Research Program on Neutron Scattering of IMSS, KEK.

\appendix

\section{Modeling the saturation of multiplication} \label{sec:SaturationModel}
Suppose that we have a single-wire drift tube with sense-wire radius, $a$, and tube radius, $b$. 
We take a cylindrical coordinate system with the $z$-axis aligned in the wire direction and
consider a multiplication process problem in which a dense bunch of ionized electrons drift to a narrow spot on the wire.
Assume that the process is sufficiently fast, i.e., the ions created by avalanche multiplication do not move away from their created position close to the wire in the time being considered.
Also, we approximate the model by assuming that every ionized electron multiplication process is instantaneous and sequential in time; an ionized electron drifting to the wire only looks at the electrostatic field around the wire, which was formed by all previous multiplications.
The space charge distribution around the sense wire at the time that $n$ ionized electrons have already multiplied is labeled $N_{n}(r)$.
From Gauss's law, the electric field, $E_{n}(r)$, is written as
\begin{eqnarray}
2\pi r \epsilon E_{n}(r) = e\sigma d\int_{a}^{r}dr'\int_{0}^{d}dz'2\pi r' N_{n}(r'),
\end{eqnarray}
where $\epsilon$, $e$, $\sigma$, and $d$ denote the dielectric constant, elementary charge, line charge density on the wire, and system dimension in the $z$-direction, respectively. The boundary condition is
\begin{eqnarray}
\int_{a}^{b}dr\,E_{n}(r) = V_{0},
\end{eqnarray}
where $V_{0}$ is the high voltage applied to the sense wire. The form of $N_{n}(r)$ is not known precisely but, if we assume that the shape of the distribution does not change drastically during the multiplication process, we can write
\begin{eqnarray}
N_{n}(r) = \left(\sum_{j=0}^{n-1}G_{j}\right)\cdot \bar{\rho}(r),
\end{eqnarray}
where $G_{j}$ is the multiplication gain for the $j$-th ionized electron and $\bar{\rho}(r)$ is the average space-charge distribution, normalized by
\begin{eqnarray}
\int_{a}^{b}dr\,\bar{\rho}(r)=1~.
\end{eqnarray}
Then, the electric field is formally expressed as
\begin{eqnarray}
E_{n}(r) &=& \frac{V_{0}}{r\ln(b/a)}+\frac{e}{\epsilon ad}\left(\sum_{j=0}^{n-1}G_{j}\right)\nonumber\\
&&\times\left[-\frac{1}{\ln(b/a)}\int_{a}^{b}dr'\,\frac{\bar{\rho}(r')}{r'} + \bar{\rho}(r)\right],
\end{eqnarray}
where the first term is the electric field with no space charges, the second term is the ``screening effect'' of the electric field due to the space charges, and the third term is the Coulomb field created by the space charges. We do not calculate the multiplication gain, $G_{n}$, for the given $E_{n}(r)$ explicitly. Alternately, we suppose a real value, $\bar{E}_{n}$, which characterizes the scale of $E_{n}(r)$. Under the assumption that $\bar{\rho}(r)$ does not change, we can define $\bar{E}_{n}$ as the value $E_{n}(r=a)$. For example
\begin{eqnarray}
\label{eq:field_equation_simple}
\bar{E}_{n} &\equiv& E_{n}(r=a) = \frac{V_{0}}{a\ln(b/a)}-\frac{eC}{\epsilon ad}\left(\sum_{j=0}^{n-1}G_{j}\right),\\
C &=& \frac{1}{\ln(b/a)}\int_{a}^{b}dr'\,\frac{\bar{\rho}(r')}{r'} - \bar{\rho}(a) = {\rm const}~~.
\end{eqnarray}

\begin{figure}[htbp]
\begin{center}
\includegraphics[width=0.6\textwidth]{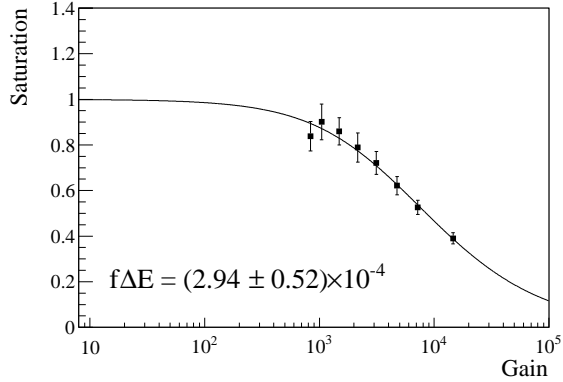}
\caption{Measurement of the TPC's saturation $s(G_{0})$ as a function of gain
for a certain energy deposit on a sense wire given by the ${\rm
{}^{241}Am}$ source injection. The vertical scale was normalized to
$s(G_{0})\rightarrow 1$ at the limit of $G_{0}\rightarrow 0$ by
fitting the data points with Eq. \ref{eq:saturation_relation} multiplied by an
arbitrary scale. The product of the saturation parameter $f$ and the
energy deposit $\bar{\Delta E}$ is obtained by the fitting.}
\label{fig:gainscan}
\end{center}
\end{figure}

Then, we employ an empirical relation between $G_{n}$ and $\bar{E}_{n}$, which is widely applicable to wire chambers
\begin{eqnarray}
\label{eq:gain_as_a_func_of_field}
G_{n} = \exp\left(A + B\bar{E}_{n}\right),
\end{eqnarray}
where $A$ and $B$ are arbitrary constants used to characterize the multiplication.
From Eq.~(\ref{eq:field_equation_simple}) and (\ref{eq:gain_as_a_func_of_field}), if we take the difference, $\bar{E}_{n}-\bar{E}_{n-1}$, we can deduce a recurrence relation for the gain such that
\begin{eqnarray}
G_{n} = G_{n-1}\exp\left(-\frac{eBC}{\epsilon ad}G_{n-1}\right)~.
\end{eqnarray}
The above recurrence relation is approximately solved in the case of $\alpha G_{0}\equiv eBC/(\epsilon ad)\cdot G_{0}\ll1$ to
\begin{eqnarray}
\frac{G_{n}}{G_{0}} = \frac{1}{1+\alpha G_{0}n}~.
\end{eqnarray}
By summing over all $N$ ionized electrons and normalizing by the total multiplication, assuming no space-charge effects, the degree of degradation of the multiplication caused by the space-charge effect is expressed as
\begin{eqnarray}
\label{eq:saturation_relation}
s(G_{0}) &\equiv& \frac{1}{NG_{0}}\left(\sum_{j=0}^{N}G_{j}\right) = \frac{\ln(1+\alpha N G_{0})}{\alpha N G_{0}},\nonumber\\
&=& \frac{\ln(1+f\Delta E G_{0})}{f\Delta E G_{0}},
\end{eqnarray}
where $\Delta E$ is the amount of energy deposit observed by the wire, which is proportional to $N$. The coefficient, $f$, is hence defined as $f\equiv \alpha N/\Delta E$. We call $f$ the {\it saturation parameter}. Eq.~(\ref{eq:saturation_relation}) indicates that $f$ is a measurable value, provided $\Delta E$ has been determined by scanning $G_{0}$ and relatively comparing the variance of $s(G_{0})$. 
We performed the measurement of $s(G_{0})$ for our TPC with
configuring a ${\rm {}^{241}Am}$ source collimated to fixed incident
angles. Figure \ref{fig:gainscan} shows an example of the scan of $s(G_{0})$ by
changing the sense wire's HV. 
The gain $G_{0}$ at a given sense wire's HV was monitored using a
${\rm {}^{55}Fe}$ 5.9 keV X-ray source.
The common absolute normalization for
the set of the HV scan and $f\bar{\Delta E}$ are fitted simultaneously
following Eq.~(\ref{eq:saturation_relation}) (the vertical scale of Figure \ref{fig:gainscan} is
already normalized using the fitted absolute scale so that $s(G_{0})$
becomes 1 at $G_{0}\rightarrow 0$). The deposit energy on the wire
$\bar{\Delta E}$ is undetermined from the measurement, and it was
estimated using Geant4 simulation by comparing the energy deposit on
different MWPC wires to extract the saturation coefficient $f$.
Once $f$ is obtained, it is possible to calculate the space-charge effect as a function of the primary gain, $G_{0}$. Eq.~(\ref{eq:saturation_relation}) does not explicitly depend on the precise structure of the single-wire drift tube (e.g., parameters such as the wire radius), and we can obtain the same equation for MWPCs; the precise structure related to the saturation is all included in the saturation parameter, $f$.

\begin{figure}[htbp]
\begin{center}
\includegraphics[width=0.6\textwidth]{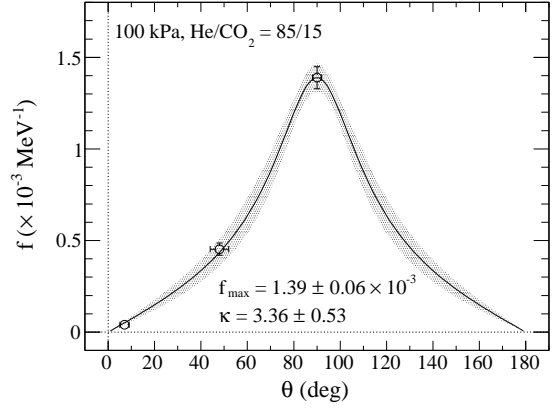}
\caption{Saturation parameter, $f$, as a function of the particle track angle, $\theta$, projected onto the $xz$-plane with respect to the wire.
The parameters, $f_{\rm max}$ and $\kappa$, are obtained from the measurements indicated by circles.
}
\label{fig:saturation}
\end{center}
\end{figure}

Let us then consider determining $f$ in the case of our TPC, for a dense track. 
We use the same coordinate system defined in Section \ref{sec:overview}.
It is obvious that $f$ depends on the effective length of the wire, which is affected by the space-charge effect. 
Provided the drifting of positive ions away from the sense wire is negligible, we can neglect the time dependence, and the system is then two-dimensional. 
Thus, $f$ should be a function of the angle of the particle track, $\theta$, projected onto the $xz$-plane with respect to the wire. 
Qualitatively, we can say that $f$ takes the largest value at $\theta=\pi/2$ and the smallest at $\theta=0$. 
Also, $f$ should be even for $\theta<\pi/2$ and $\theta>\pi/2$, and it is smooth and at $\theta=\pi/2$. 
The simplest form of $f(\theta)$ will be written as a function of 
\begin{eqnarray}
f(\theta) = \frac{f_{\rm max}}{\sqrt{1+(\kappa\cot\theta)^{2}}},
\end{eqnarray}
where $f_{\rm max}$ is the maximum saturation parameter at $\theta=\pi/2$ and $\kappa$ is the strength of the $\theta$-dependence. 
These parameters can be obtained from the actual measurement by using a $^{241}$Am source, as shown in Figure \ref{fig:saturation}.
Mathematically, the energy deposit density on the wire diverges at $\theta=\pi/2$ 
but, in reality, it remains finite due to multiplication dispersion in the $z$-direction, as well as the diffusion of ionized electrons before multiplication.
At the $\theta\rightarrow 0$ limit, $f\rightarrow 0$ and $s(G_{0})\rightarrow1$ and no space-charge effect exists.

\newpage
\bibliographystyle{elsarticle-num}
\bibliographystyle{unsrt}
\bibliography{main}


\end{document}